# ROSAT OBSERVATIONS OF BLAZARS FROM A POLARIZED RADIO SELECTED SAMPLE.

L. Maraschi[1,2], G. Fossati[2,3], G. Tagliaferri[4], and A. Treves[3]

[1] Dipartimento di Fisica, Università di Genova, via Dodecaneso 33, 16146 Genova, Italy

[2] Dipartimento di Fisica, Università di Milano, via Celoria 16, 20133 Milano, Italy

[3] SISSA/ISAS, via Beirut 2/4, 34014 Trieste, Italy

[4] Osservatorio Astronomico di Brera, via Brera 28, 20121 Milano, Italy

# Abstract.

We have observed with *ROSAT* six blazars from the sample of 31 polarized objects constructed by Impey and Tapia (1988). Results of the spectral analysis in the *ROSAT* band of three objects, PKS 0301−243, PKS 1055+018 and PG 1424+240 are presented here and compared with the three previously published objects. Using a simple power law model with free absorption a wide range of spectral slopes is found, from 0.2 to 2.2 in the energy spectral index $\alpha_X$. Objects with emission lines have flatter X–ray spectra than lineless objects. Indication of spectral curvature within the X–ray band is found in the sense of a concavity (soft excess) for the flattest spectrum and a convexity (soft deficiency) for the steepest one. The X–ray spectral shape is discussed in relation to the overall energy distribution for the six objects. The results are compared with other samples of blazars. The X–ray spectral index, $\alpha_X$, is found to be anticorrelated with the broad band spectral index $\alpha_{RO}$ and with bolometric luminosity and redshift. It is suggested that the different spectra of quasar–like and BL Lac–like blazars represent extremes in a continuous distribution rather than two distinct populations.

*subject headings*: BL Lacertae objects: individual (PKS 0301–243, PG 1424+240) — galaxies: active — quasars: individual (PKS 1055+018) — X-rays: galaxies.

# 1. Introduction

In the spirit of unified models for active galactic nuclei, it is interesting to focus on the relation between BL Lac objects and Flat Spectrum Radio Quasars (FSRQ), which include OVVs (Optically Violent Variables) and HPQs (High Polarization Quasars). BL Lacs and FSRQ show a close similarity in the properties of their non thermal continuum such as the flatness of the radio spectrum, the high polarization and variability at all frequencies (Impey & Tapia, 1990; Gear et al. 1994; Bregman, 1990, 1994). They are however distinguished by the strength of the emission lines, (a conventional threshold is put at an equivalent width of 5 Å, e.g. Stocke et al. 1989), by the luminosity and redshift distribution (e.g. Ghisellini et al. 1993), and by the morphology of the extended radio emission, being mostly of FR I and FR II type (Fanaroff & Riley, 1974) for BL Lacs and FSRQ respectively.

For both classes of objects the non thermal emission is thought to be produced in a relativistic jet closely aligned with the line of sight and therefore amplified by relativistic beaming. Unification schemes have been proposed whereby FR I and FR II radio galaxies represent the unbeamed version (or parent population) of BL Lacs and FSRQs respectively (e.g. Urry, Padovani and Stickel 1991; Padovani, 1992a).

The properties of BL Lac objects and FSRQ could also fit into a generalized unification framework, whereby the relativistic jet phænomenon occurs in a similar way, but with a range in the intrinsic power of the jet, in galaxies with a range of other activities, possibly related to accretion power and/or the ambient medium (Maraschi & Rovetti, 1994). In order to make progress in this direction it is important to compare the energy distributions of the different classes of objects and to account for possible systematic differences between them.

Originally, the knowledge of the spectral properties of BL Lac objects developed within two distinct sub-classes of sources affected by different strong observational biases. The first group are the so called *Radio-selected BL Lacs* (RBL), objects discovered from the study of radio flux limited complete samples, while the second group, *X-ray selected BL Lacs* (XBL), derives from X-ray flux limited samples. The typical overall energy distributions of the two groups are systematically different. The XBL show a smooth radio to UV spectrum with a broad peak in the IR-UV range, and a steep soft X-ray tail while the RBL have a distribution peaked in the FIR band and a flatter X-ray spectrum (Ghisellini et al. 1986; Maraschi et al. 1986; Padovani, 1992b; Gear, 1994; Maraschi, Ghisellini & Celotti, 1994; Pian et al. 1994). They also differ in their optical variability and polarization properties, with XBLs showing less extreme behaviour (Jannuzi et al. 1994).

The Broad Band Energy Distributions (BBED) of FSRQs are similar to those of RBLs, having typically a peak in the far-infrared/submillimeter range and a flat X-ray spectrum (Ghisellini et al. 1986; Impey & Neugebauer, 1988; Worrall & Wilkes, 1990; Pian & Treves,

1993; Brunner et al. 1994).

Previous X–ray observations with *Einstein* (Worrall & Wilkes, 1990) and *EXOSAT* (Sambruna et al. 1994) have shown that the X–ray spectra of the XBLs are usually steep and convex, while those of RBL and FSRQ are flatter, suggesting the general picture that this could be due to a different contribution from two emission components associated with the synchrotron and inverse Compton mechanisms respectively (e.g. Maraschi, Ghisellini & Celotti 1994). In this regard the *ROSAT* PSPC band is particularly useful since it can be sensitive to both spectral components.

Here we discuss new data from the *ROSAT* satellite on a sample of objects selected by Impey and Tapia (1988, 1990) on the basis of their optical polarization and disregarding the presence or absence of emission lines. The resulting sample being unbiased with respect to line emission allows a more objective study of blazars considered as a single class. Since the optical counterparts were required to be stellar, the sample does not contain nearby objects for which the surrounding galaxy is resolved on the Palomar Sky Survey plates.

The 31 blazars of the Impey and Tapia sample (1988) derive from the quasar–like sources of the complete radio surveys, homogenized by Kühr et al. (1981), with threshold at 2 and 1.5 Jy at 5 GHz, to which 51 sources of the 3CR catalog stronger than 9 Jy at 178 MHz have been added. The criterion for the inclusion in the Impey and Tapia sample is the existence of an optical polarization measurement with $p > 2.5$ %.

Our *ROSAT* targets were selected taking into account X–ray brightness from *Einstein* when available, the optical magnitude, and the hydrogen column density. Since the program was approved in C category, satellite constraints concurred in the final target selection. Data were obtained for six objects. Results on three of them were reported in Treves et al. (1993) and Maraschi et al. (1994). While obviously the sample is not complete, it happens to contain three lineless (BL Lac) objects and three quasar–like objects, which may be representative of the original sample of radio bright, flat spectrum ($\alpha_R < 0.5$) blazars. Hereafter the spectral index will be defined by $F_\nu \propto \nu^{-\alpha}$, where $F_\nu$ is the spectral energy flux (e.g. Jansky). All are highly polarized, $p > 4.6$ %.

Section II presents the spectral analysis of the *ROSAT* data of three objects and summarizes the results for the subsample of 6 objects. In this section the spectral index of the photon flux is used as defined by $N_E \propto E^{-\Gamma}$, where $N_E$ is the photon flux (e.g. photons cm$^{-2}$ sec$^{-1}$ keV$^{-1}$.

We then discuss the spectral shapes in the X–ray band in relation with the overall energy distributions (section III). The results are compared with other samples of radio and X–ray selected blazars observed with *ROSAT* (*e.g.* Brunner et al. 1994; Brinkmann & Siebert, 1994) (Section IV).

## 2. *ROSAT* observations and analysis

PKS 0301–243, PKS 1055+018 and PG 1424+240 were observed with *ROSAT* in 1993. For PKS 0301–243 and PG 1424+240, spectroscopy has never revealed the presence of emission lines in the optical spectrum. Featureless spectra were reported by Condon et al. (1977) in the course of the optical identification program of the Parkes radio sources, and recently by Véron–Cetty and Véron (1993) for PKS 0301–243 and Thompson et al. (1990) for PG 1424+240. These sources can therefore be classified as BL Lac objects and their redshift is unknown. For PKS 0301–243 a redshift estimate ($z \simeq 0.2$) was derived by Falomo et al. (1993a) on the basis of the best decomposition of the simultaneous IR–optical spectrum in terms of a power law plus a contribution of the host galaxy.

It is interesting to note that PG 1424+240 was misclassified as a DC white dwarf in the Palomar Green survey, even if a radio source was known at this position (Condon et al. 1977). The optical polarization study by Impey and Tapia (1988) made no reference to the PG source. PG 1424+240 was recently reclassified as a BL Lac object (together with other two objects) in the course of a study of PG white dwarfs detected in the *ROSAT* All Sky Survey (RASS) (Fleming et al. 1993). Therefore this source satisfies selection criteria in widely different frequency bands: radio (Kühr et al. 1981) optical (PG survey) and X–rays (*ROSAT* All Sky Survey).

For PKS 1055+018 there is a redshift determination ($z = 0.888$, Medd et al. 1972, Lynds & Wills, 1972) based on the CIII$\lambda$1909 emission line, which is reported as "prominent". Therefore this object can be classified as a FSRQ.

The relevant information about the three *ROSAT* observations is reported in Table 1. The extraction of the events from the PSPC photon list and the preparation of the binned spectra were performed with the MIDAS/EXSAS software (Zimmermann et al. 1993). The source counts were extracted using circular areas centered on the target positions with radii between 3 and 3.5 arcmin. The background contribution was evaluated on surrounding circular annuli with external radii 9 − 12 except in the case of PKS 1055+018 where diffuse emission close to the target is present. A search through NED (NASA Extragalactic Database), reveals in fact a positional coincidence with the Abell cluster A1139 at $z = 0.038$ (Abell et al. 1989) centered at $\sim 6$ arcmin from the PKS 1055+018 and extending for 70 arcmin. The cluster is obviously foreground relative to the blazar. The observed extended emission is probably related to emission from the intracluster gas in the deeper part of the cluster potential well. To properly perform the extraction of the source and background counts we ignored a sector of the image including most of the extended emission region.

The collected counts and the deduced source count rates are reported in Table 1. Standard corrections for vignetting, dead time and effective area were applied. No variability is apparent from inspection of the X–ray light curves, beyond that likely due to the intrinsic wobbling of the satellite.

In each case spectra were produced from the *ROSAT* instrumental binning, ignoring the first 10 channels, in such a way that each rebinned channel has S/N ratio larger than 5. This yielded 26 bins for PKS 0301–243, 12 bins for PKS 1055+018 and 51 bins for PG 1424+240. In each case the rebinning did not degrade the intrinsic energy resolution of the instrument. The spectral analysis was then performed with the XSPEC package (Shafer et al. 1991). We analysed the spectra both with the matrix version provided with the data and with the most recent release, finding no significant difference in the results.

We used a single power law model with absorption by neutral gas with standard cosmic abundances ($\tau = n_H \sigma(E)$, cross section $\sigma(E)$ from Morrison & McCammon, 1983), considering $n_H$ as a free parameter. The best fit was obtained through $\chi^2$ minimization. The results are reported in Table 2. Quoted errors correspond to 90 % confidence intervals for two interesting parameters. The confidence contours for $\Gamma$ and $n_H$ are shown in Figures 1a,b,c. The vertical dotted lines represent the galactic $n_H$ values derived from Dickey and Lockman (1990). A band of uncertainty of $\pm 0.3 \times 10^{20}$ cm$^{-2}$ as estimated from a comparison of measurements with different angular resolution (Lockman et al. 1986, and Stark et al. 1992) is also shown in Figure 1.

For PKS 0301–243 the best fit column is in good agreement with the galactic value and the fit is acceptable in terms of reduced $\chi^2$.

For PKS 1055+018 the fit with free $n_H$ is good but the best fit $n_H$ is significantly lower than the galactic value. In fact, a fit with $n_H$ fixed at the galactic value gives a rather high value of $\chi^2$ (probability of $9 \times 10^{-2}$). The F-test (Bevington, 1969) shows that the improvement of the fit with free $n_H$ is significant at the 0.987 level, not probatory but indicating the presence of excess emission in the soft band.

In the case of PG 1424+240 the fit with free $n_H$ is acceptable but in this case the derived value of the column density is substantially higher than the galactic one. A fit with fixed absorption turns out to be poor, with a $\chi^2$ probability of $8.1 \times 10^{-5}$. The F-test shows that the improvement for free $n_H$ is highly significant (probability of $3 \times 10^{-8}$ for the null hypothesis).

One cannot entirely rule out the possibility of anomalous fluctuations of the Galactic absorption along the line of sight, nor, in the case of PG 1424+240, the possibility of local absorption at the source. However it is interesting to consider that the indicated soft excess or deficiency may be intrinsic to the emitted spectrum. In order to quantify deviations from a single power law we therefore fitted the data for PKS 1055+018 and PG 1424+240 with a broken power law model. $n_H$ was kept fixed at the galactic value.

The adopted spectral model is:

$$N(E) = \begin{cases} K \left(\dfrac{E}{E_0}\right)^{-\Gamma_1} e^{-n_H \sigma(E)} & E \leq E_{break} \\ K \left(\dfrac{E_{break}}{E_0}\right)^{\Gamma_2 - \Gamma_1} \left(\dfrac{E}{E_0}\right)^{-\Gamma_2} e^{-n_H \sigma(E)} & E \geq E_{break} \end{cases} \quad (1)$$

Results are reported in Table 2. The errors quoted for PG 1424+240 correspond to 68 % confidence intervals for three interesting parameters. One can see that in terms of $\chi_\nu^2$ the fits are both formally acceptable. The F–test applied for the significance of the improvement between the simple and the broken power law results, gives a probability against the null hypothesis of 0.985 for PG 1424+240 and only of 0.75 for PKS 1055+018. In the case of PKS 1055+018, the sparseness of the data and the consequent small number of bins make it difficult to firmly discriminate between the two models, and moreover make it impossible to obtain combined errors for the three interesting parameters. Thus we can only get a weak indication that the $ROSAT$ spectrum of PKS 1055+018 could be well (better) described with a broken power law with a suggested concave shape. The case of PG 1424+240 is on the contrary quite clear, with a resulting convex spectrum. The lower limit error on $\Gamma_1$ is strongly affected by the finiteness of the $ROSAT$ band, even though for this source the number of bins is high and hence the energy partition quite fine. Considering then as statistical error on $\Gamma_1$ its upper error interval we obtain for the spectral break $\Delta\Gamma = 1.45 \pm 0.55$ (1 $\sigma$).

In Table 2 we also report results for PKS 0537–441, PKS 1034–203, PKS 1335–127 for which a similar kind of analysis on $ROSAT$ data was performed (Treves et al. 1993; Maraschi et al. 1994). While for two sources the single power law model gave an adequate representation of the data, for PKS 1034–293 a significant soft excess was found, which could be described by a broken power law with concave curvature. A power law model with an absorption feature was also fitted to the spectrum of PKS 1034–293 and gave results statistically equivalent to the broken power law. However the depth and position of the absorption feature are difficult to interpret (Maraschi et al. , 1994). Therefore, pending also some clarification on possible gain shifts in the instrument, this alterntive was not further discussed.

Comparing now the six sources in Table 2, we note that there is a wide spread in the spectral indices derived with the single power law model, $(\alpha_X)_{max} = 2.17^{+0.28}_{-0.27}$, $(\alpha_X)_{min} = 0.23^{+0.54}_{-0.50}$ significantly larger than the typical error on individual values. Moreover, the source with the flattest X–ray spectrum shows evidence of a concave shape and the source with the steepest X–ray spectrum shows evidence of a convex shape. The fact that the sources which show deviations from a single power law model in their $ROSAT$ spectra are also those showing extreme values of the X–ray spectral index supports the hypothesis that the inferred "curvature" reflects a property of the continuum rather than absorption or instrumental effects.

## 3. Broad Band Energy Distributions

Broad Band Energy Distributions were constructed for the three objects discussed above through a careful search in literature. The selection of data was based on the homogeneity and simultaneity, when available, of measurements in adjacent bands (Radio and Millimeter, IR and Optical–UV). The radio, optical and X–ray fluxes for the six sources, the spectral indices in those bands and the computed "two band" indices $\alpha_{RO}, \alpha_{OX}, \alpha_{RX}$, along with references are reported in Table 3. The two band effective spectral indices are defined as usual, $\alpha_{12} = -\log(F_1/F_2)/\log(\nu_1/\nu_2)$ and are taken between the frequencies corresponding to 5 GHz, 2500 Å and 2 keV. The 2 keV flux values were deduced from the 1 keV fluxes using the best fit spectral indices. The flux at 2500 Å when not available was obtained extrapolating from the optical.

The resulting BBEDs in the $\nu F_\nu$ representation are shown in Figure 2, together with those of the other three sources for which the $ROSAT$ spectrum was discussed previously. The fluxes were scaled (two decades per source) to permit a common representation and the sources were ordered according to the radio to optical spectral index $\alpha_{RO}$. The results of the analysis of the $ROSAT$ data are reported in the form of a symbol corresponding to the 1 keV flux and of three solid lines, representing the best fit slope and the 90 % confidence limits for the single power law fits.

The BBED of PKS 0301–243 and PG 1424+240 are remarkably similar with a broad maximum between $10^{14}$ and $10^{16}$ Hz. The IR–UV spectra are rather flat: for PKS 0301–243 a constant value $\alpha_{IR-OP-UV} = 0.84$ was deduced from a power law fit to the data in the whole range allowing for a host galaxy contribution at $z = 0.2$ (Falomo et al. 1993a); for PKS 1424+240 $\alpha_{IR-OP} = 1.02$, was deduced from the data of Mead et al. 1990). The soft X–ray spectra can be smoothly connected with the optical–UV continuum, implying a continuous steepening of the energy distribution from optical to X–ray frequencies. It is interesting to note that the convex spectral shape of PG 1424+240 in the $ROSAT$ band, indicated by the broken power law fit, is consistent with the trend of the overall energy distribution. This suggests that the X–ray emission is an extension of the lower energy spectrum, with a continuously increasing downward curvature.

On the other hand, the BBED of PKS 1055+018 is remarkably different from the two described above: it shows a steeply declining IR–optical spectrum ($\alpha_{IR-OP} = 1.6$, Falomo et al. 1993b) beyond a peak in the power per logarithmic bandwidth occurring presumably in the $10^{12} - 10^{14}$ Hz range. The X–ray flux exceeds by two orders of magnitude an extrapolation of the optical spectrum and the X–ray spectrum is hard, $\alpha_X < 1 < \alpha_{IROP}$, implying an upturn in the energy distribution at high energy. Again the suggested concavity in the $ROSAT$ band matches well the overall energy distribution. Only the excess emission in the soft X–ray band could perhaps be related to the ultraviolet continuum.

We recall that in Figure 2 the sources are ordered from bottom to top according to increasing values of $\alpha_{RO}$, *i.e.* a parameter independent of the X–ray measurements. The six sources show a remarkable regularity in their BBEDs: the spectral behaviour in the X–ray band appears to be strongly correlated with the shape of the Radio–Optical continuum, in the sense that the lower is the value of the effective index $\alpha_{RO}$, the steeper is the $ROSAT$ X–ray slope. When the X–ray spectrum is steep the X–ray emission can be smoothly connected to the lower frequency continuum (PKS 0301–243, PKS 1424+240), while flat X–ray spectra occur in objects with a steep optical continuum.

There are transition objects (PG 1335–127 and PKS 0537–441), where the X–ray optical connection is ambiguous. On this regard it is interesting to note that PKS 0537–441 was classified as a BL Lac by Stickel et al. (1991,1993) but a relatively strong MgII line is sometimes observed. For the same object, comparing *Einstein*, EXOSAT and $ROSAT$ data Treves et al. (1993) inferred a variability of the X–ray spectral shape, with a brighter state corresponding to a softer spectrum.

## 4. Comparison with other Samples

We can compare our results with the main existing samples of blazars observed with $ROSAT$.

The X–ray spectra of BL Lacs which were bright in the $ROSAT$ All Sky Survey were analysed by Brinkmann and Siebert (1994). These comprise 19 objects, of which 2 can be considered radio loud (RBL) and 17 radio weak (XBL). A complete sample of FSRQ (8 objects) was studied by Brunner et al. (1994). The spectral index distributions of the two samples are compared with ours in Figure 3. Clearly the FSRQs of Brunner et al. (1994) have flatter X–ray spectra than the X–ray bright BL Lac objects. The mean spectral indices for the two samples are $\langle \alpha_X \rangle_{X-loud} = 1.56 \pm 0.43$, $\langle \alpha_X \rangle_{FSRQ} = 0.59 \pm 0.18$ respectively.

It is notable that the spectral indices found for our 6 objects span a wide range of values, from 0.2 to 2.2 in the energy spectral index $\alpha_X$, covering almost the full range of the other two samples. This spread is real since the "typical" error for the individual index is $\sim \pm 0.5$. The mean spectral index for our six sources is $\langle \alpha_X \rangle = 1.18 \pm 0.64$, intermediate between the two above and with larger dispersion.

Analysis of the $ROSAT$ data for the two main complete samples of BL Lacs, those of radio–selected BL Lacs by Stickel et al. (1991) and of X–ray–selected BL lacs from the EMSS (Morris et al. 1991) is in progress. The average values are $\langle \alpha_X \rangle_{Stickel} = 1.10 \pm 0.43$ (Sambruna et al. 1993) and $\langle \alpha_X \rangle_{EMSS} = 1.20 \pm 0.46$ (Perlman et al. 1994). Both agree with the average $\alpha_X$ for our six objects. At present we believe that this equality of the average spectral shapes in the ROSAT band is the result of a chance coincidence. In fact, below the ROSAT band, at optical-UV frequencies, RBL spectra are on average steeper than XBL, while the opposite is true above the ROSAT band, in the medium X-ray range (Ghisellini et al. 1986, Worral and Wilkes 1990, Sambruna et al. 1994). A more detailed discussion should await completion of the work on the latter samples.

It is important to examine whether the correlation between the radio to optical energy distribution and X-ray spectral shape apparent from Figure 2 also holds for the other available samples. Figure 4a shows the $ROSAT$ X-ray spectral index plotted against $\alpha_{RO}$ for the six sources discussed here and for the samples studied by Brinkmann and Siebert (1994) and by Brunner et al. (1994). A clear anticorrelation is present if we consider all the sources together, in the sense of flatter X-ray spectra for steeper values of $\alpha_{RO}$.

We tested the significance of the correlation with the Spearman rank-order test and with Kendall's $\tau$ test (Press et al. 1994) finding in both cases $p > 99.99$ %. Kendall's correlation coefficients and associated probabilities are reported in Table 4.

The difference $\triangle \alpha = (\alpha_{OX} - \alpha_X)$, indicates how the X-ray spectral shape connects with the optical continuum: null or negative values imply a straight or convex connection, while positive values correspond to a concave connection indicating the existence of a separate hard X-ray component. In Figure 4b $\triangle \alpha$ is plotted against $\alpha_{RO}$. As expected from Figure 2 this curvature indicator is correlated with $\alpha_{RO}$. In fact the significance of the latter correlation is higher than the previous one (see Table 4).

Although $ROSAT$ X-ray data on a complete sample of RBL objects are not yet available, one can get some information on their position in the diagrams by means of the published average values. RBLs are characterized by a $\langle \alpha_{RO} \rangle \simeq 0.6$ (Padovani, 1992b, for the Stickel complete sample) with a small dispersion. Using the average spectral index reported by Sambruna et al. (1993) for the Stickel sample, $\langle \alpha_X \rangle = 1.1 \pm 0.4$, one can anticipate that in Figure 4a RBLs would likely occupy the region between XBLs and FSRQ, in agreement with the correlation noted above.

One can see from Figure 4 that there is a continuous range of properties from BL Lacs to FSRQs. In particular the arrangement of the six sources discussed here seems to stress this continuity covering almost the entire range of parameter values and connecting the two populations. This is likely to be the result of the selection procedure that originated the sample from which the six were taken which was "objective" with regard to line emission. It is also worth noting that the two lineless objects, although they are in some way radio-selected, fall in a region typical of XBLs.

From Figures 4a and 4b a picture emerges where XBL, RBL and FSRQ have continuous, albeit different, properties, and it is interesting to examine what physical or astrophysical parameter could govern this behaviour. Here we can consider redshift and /or luminosity. In fact it is well known that XBLs are closer and less luminous than FSRQs (e.g. Padovani 1992a).

Considering only the 26 objects with known redshift we looked for correlations of $\alpha_X$, $\triangle\alpha = (\alpha_{OX} - \alpha_X)$ and $\alpha_{RO}$, with $z$ and/or optical luminosity which is close to the peak in the distribution of the power per decade (see Figure 2). The optical luminosity was approximated using the monochromatic emitted power $L_O = 4\pi d_L^2 F_O$ multiplied by the frequency $\nu_O$ ($H_0 = 50$ Km sec$^{-1}$ Mpc$^{-1}$, $q_0 = 0.5$). We found significant correlations ($p > 99.9$ %) of $\alpha_{RO}$, $\alpha_X$, and $\triangle\alpha$ with $z$ and weaker ones ($p > 95$ %) with $\nu_O L_O$. A partial correlation test (Padovani, 1992b), subtracting the effect of $\nu_O L_O$ or $z$ or both, gives a net correlation between $\alpha_X$ and $\alpha_{RO}$, still significant at 99.32 %, and between $\triangle\alpha$ and $\alpha_{RO}$ significant at 99.86 % (see Table 4). We therefore conclude that the correlation between the X–ray spectral index and the broad band index $\alpha_{RO}$ is genuine, i.e does not result from a common dependence of the two indices on either the observed luminosity or the redshift. The fact that $\alpha_{RO}$ is strongly correlated with redshift, in the sense of steeper $\alpha_{RO}$ at larger red-shift, suggests that the systematic changes in the full broad band energy distribution may be associated with cosmological evolution. Selection effects related with optical identification of faint sources should favor the opposite trend. However one can not entirely exclude that other, unknown biases may be responsible for the observed effect.

## 5. Summary and conclusions

We have obtained *ROSAT* data on 6 blazars drawn from a sample of flat spectrum radio sources with star–like optical counterparts and high optical polarization, disregarding line emission properties. Three of the objects happen to be of BL Lac type (lineless) while three are quasar–like (with broad emission lines). We find widely different X–ray spectra, with quasar–like objects having substantially flatter spectra than BL Lac objects. We also find a strong correlation of the X–ray spectral index with the radio–optical broad band spectral index, which in turn correlates with redshift and, more weakly, with observed luminosity.

Objects with flat overall radio to optical energy distributions have steep X–ray spectra smoothly connecting with the lower frequency continuum through a progressive spectral steepening. This suggests a single emission mechanism, i.e. synchrotron radiation, from radio to X–ray frequencies. The progressive steepening could be due to the energy spectra of the emitting particles being "curved" or to inhomogeneity of the emission region (e.g. Maraschi, Ghisellini & Celotti 1994). Objects with steep radio to optical energy distribu-tions (high $\alpha_{RO}$) have flat X–ray spectra, suggesting that the emission mechanism in the X–ray band is different from synchrotron, and is most likely inverse Compton emission.

Despite the diversity mentioned above, combining our data with *ROSAT* spectral data for X–ray bright BL Lacs and for FSRQs, all the spectral properties point to a continuous sequence rather than to two distinct populations. Moreover, the average values of the spectral parameters for radio–selected BL Lacs of the complete Stickel sample seem to be intermediate between X–ray bright BL Lacs and FSRQs.

We therefore conclude that blazars can be considered as a single population in which the shape of the broad band continuum from radio to optical *and the consequent shape of the X–ray spectrum* are related to some physical quantity which varies continuously across the population. The same quantity should also be responsible for the strength of the emission lines since FSRQs represent one extreme within the global population. From the information discussed in this paper it appears that the effect may be associated with redshift and therefore could be connected to cosmological evolution.

ACKNOWLEDGEMENTS

This research has made use of the NASA/IPAC Extragalactic Database (NED) which is operated by the Jet Propulsion Laboratory, Caltech, under contract with the National Aeronautics and Space Administration.

# Figure Captions.

FIG. 1: $\Gamma/n_H$ contours at 68, 90, 99 % confidence for two interesting parameters. The vertical lines represent the galactic $n_H$ value with an uncertainty of $\pm\, 0.3 \times 10^{20}$ cm$^{-2}$: (a) PKS 0301–243; (b) PKS 1055+018; (c) PG 1424+240.

FIG. 2: The Radio/X–ray Broad Band Energy Distributions for the whole sample in the $\nu F_\nu$ form. Sources are ordered from bottom to top according to their $\alpha_{RO}$ value. The ordinate scale is correct for PG 1424+240. The fluxes of the other sources have been shifted upwards by two decades each. The short dotted lines along the left scale display the $10^{-14}$ level for each source. Radio to submillimeter data are taken from the Parkes catalog, Quiniento et al. (1988), White & Becker (1992), Gear et al. (1994), Steppe et al. (1988, 1992). The data in the IRAS band are from Impey & Neugebauer (1988), with arrows representing upper limits. Infrared, optical and UV data are from Allen et al. (1982), Impey et al. (1984), Adam (1985), Lépine et al. (1985), Impey & Tapia (1988), Mead et al. (1990), Bersanelli et al. (1992), Falomo et al. (1993a, 1993b), Stickel et al. (1993). The solid lines represent the power law slopes for the best fit and for the 90 % confidence values of the X-ray slope as derived from *ROSAT* data.

FIG. 3: Distribution of energy spectral indices from *ROSAT* data on the six blazars discussed here (striped area) compared with (a) BL Lac objects in the All Sky Survey (Brinkmann and Siebert 1994), (b) pointed observations of FSRQs (Brunner et al. 1994).

FIG. 4: The broad band spectral index $\alpha_{RO}$ is plotted against (a) the *ROSAT* X–ray spectral index $\alpha_X$ (b) the spectral "curvature" $\alpha_{OX} - \alpha_X$. The six blazars discussed here are represented as filled triangles, the FSRQ from Brunner et al. (1994) as stars, and the X–ray loud BL Lac from Brinkmann and Siebert (1994) with open squares.

TABLE 1

ROSAT Observations.

|  | PKS 0301–243 | PKS 1055+018 | PG 1424+240 |
|---|---|---|---|
| Observation date: | 1993 – Aug – 10 | 1993 – Jun – 05 | 1993 – Jul – 11/12 |
| Observation span (UT): | 04:31:25.000 | 04:03:30.000 | 20:02:57.000 |
|  | 06:07:42.000 | 20:41:46.000 | 04:33:43.000 |
| Effective exposure (*sec*): | 3412 | 4324 | 3065 |
| S+B counts[1]: | $1092 \pm 33$ | $562 \pm 24$ | $2013 \pm 45$ |
| Net counts[2]: | $1001 \pm 33$ | $469 \pm 24$ | $1879 \pm 45$ |
| Average count rate (sec$^{-1}$): | $0.293 \pm 0.010$ | $0.108 \pm 0.005$ | $0.613 \pm 0.014$ |

NOTES: - (1) - counts collected in the source extraction area; - (2) - source net counts, computed from the S+B counts after subtraction of the relevant background contribution, obtained by scaling the total background counts on an area equal to that of the S+B extraction.

TABLE 2

Summary of $ROSAT$ data analysis.

| Object | class | $z$ | count rate (cts/sec) | $n_H^{gal}$ ($10^{20}\,cm^{-2}$) | single power law, free $n_H$ | | | | broken power law, fixed $n_H$ | | | |
|---|---|---|---|---|---|---|---|---|---|---|---|---|
| | | | | | $n_H$ | $\Gamma$ | $F_{1keV}$ ($\mu Jy$) | $\chi_\nu^2/dof$ | $\Gamma_1$ | $\Gamma_2$ | $E_b$ ($keV$) | $\chi_\nu^2/dof$ |
| PKS 0301−243 | BL Lac | ... | $0.293 \pm 0.010$ | 1.74 | $1.46^{+0.73}_{-0.58}$ | $2.68^{+0.33}_{-0.30}$ | $0.26 \pm 0.04$ | 0.955/23 | ... | ... | ... | ... |
| PKS 0537−441 | BL Lac | 0.894 | $0.345 \pm 0.067$ | 4.00 | $3.0^{+1.3}_{-1.3}$ | $2.1^{+0.4}_{-0.4}$ | $0.79 \pm 0.05$ | 1.06/14 | ... | ... | ... | ... |
| PKS 1034−293 | FSRQ | 0.312 | $0.090 \pm 0.005$ | 4.74 | $1.31^{+1.68}_{-1.23}$ | $1.23^{+0.54}_{-0.50}$ | $0.23 \pm 0.02$ | 0.998/7 | $4.00^{+15.10}_{-1.39}$ | $1.36^{+0.45}_{-0.44}$ | $0.45^{+0.44}_{-0.20}$ | 0.893/7 |
| PKS 1055+018 | FSRQ | 0.888 | $0.108 \pm 0.005$ | 3.96 | $1.74^{+1.59}_{-1.23}$ | $1.63^{+0.52}_{-0.49}$ | $0.25 \pm 0.04$ | 0.876/9 | 2.70 | 1.69 | 0.67 | 1.030/8 |
| PKS 1335−127 | FSRQ | 0.539 | $0.147 \pm 0.006$ | 6.02 | $6.60^{+3.00}_{-2.23}$ | $2.29^{+0.55}_{-0.51}$ | $0.50 \pm 0.04$ | 1.012/12 | ... | ... | ... | ... |
| PG 1424+240 | BL Lac | ... | $0.613 \pm 0.014$ | 2.75 | $5.24^{+0.89}_{-0.85}$ | $3.17^{+0.28}_{-0.27}$ | $1.22 \pm 0.10$ | 1.013/48 | $1.62^{+0.49}_{-4.00}$ | $3.07^{+0.26}_{-0.21}$ | $0.51^{+0.33}_{-0.22}$ | 1.012/47 |

NOTES: • quoted errors for the single power law fit are 90 % confidence intervals for 2 interesting parameters. • quoted errors for the broken power law case are 68 % confidence intervals for three interesting parameters.

TABLE 3

Summary of Broad Band Properties.

| Object | $p_{min}/p_{max}$ (1) | $F_R$ (2) | $F_{Opt}$ (3) | $F_X$ (4) | $\alpha_R$ (5) | $\alpha_{IROP}$ (6) | $\alpha_X$ (7) | $\alpha_{RO}$ (8) | $\alpha_{OX}$ (9) | $\alpha_{RX}$ (10) |
|---|---|---|---|---|---|---|---|---|---|---|
| PKS 0301–243 | 10.17/10.97 | 0.39 | $1.07^a$ | 0.26 | 0.47 | $0.84^b$ | 1.68 | 0.53 | 1.46 | 0.80 |
| PKS 0537–441 | 10.09/19.50 | 3.93 | $0.50/1.38^a$ | 0.79 | −0.04 | $1.31/1.67^a$ | 1.10 | 0.67 | 1.21 | 0.88 |
| PKS 1034–293 | 6.56/8.00 | 1.51 | $0.92^c$ | 0.23 | −0.20 | $1.04/1.53^d$ | 0.23 | 0.71 | 1.27 | 0.86 |
| PKS 1055+018 | 3.88/5.96 | 3.47 | $0.34^a$ | 0.25 | −0.28 | $1.58^a$ | 0.63 | 0.86 | 1.19 | 0.92 |
| PKS 1335–127 | 3.76/11.52 | 2.21 | $1.26^e$ | 0.50 | −0.19 | ... | 1.29 | 0.71 | 1.20 | 0.88 |
| PG 1424+240 | 0.77/5.28 | 0.31 | $3.48^f$ | 1.22 | 0.20 | $1.02^f$ | 2.17 | 0.43 | 1.39 | 0.76 |

Notes: - (1) - minimum and maximum optical polarization from Impey & Tapia (1988,1990) and Fleming *et al.*, 1993; - (2) - average flux at 5 $GHz$ in $Jy$; - (3) - flux at 5500 Å in $mJy$; - (4) - flux at 1 $keV$ in $\mu Jy$ from $ROSAT$ analysis; - (5) - spectral index between average fluxes at 2.7 $GHz$ and 5 $GHz$; - (6) - best fit spectral index from the cited references for PKS 0301–243, PKS 0537–441, and PKS 1055+018; for the other sources, spectral index estimated between 2.2 $\mu m$ and 3600 Å; - (7) - spectral index from $ROSAT$ analysis; - (8) - effective spectral index between 5 $GHz$ and 2500 Å; - (9) - effective spectral index between 2500 Å and 2 $keV$; - (10) - effective spectral index between 5 $GHz$ and 2 $keV$.

References: - (a) - Falomo *et al.*, 1993b; - (b) - Falomo *et al.*, 1993a; - (c) - Adam, 1985; - (d) - Impey & Brand, 1981; Allen, Ward & Hyland, 1982; Lépine, Braz & Epchtein, 1985; - (e) - Impey & Tapia, 1988; - (f) - Mead *et al.*, 1990.

TABLE 4

Correlation Coefficients (*Kendall's* $\tau$) and Probability.

| | $\alpha_{RO}$ | $\nu_o L_o$ | $z$ | $\alpha_{RO} - (\nu_o L_o)$ | $\alpha_{RO} - (z)$ | $\alpha_{RO} - (z, \nu_o L_o)$ |
|---|---|---|---|---|---|---|
| $\alpha_{RO}$ | ... | $\tau = 0.368$ | $\tau = 0.527$ | ... | ... | ... |
| | ... | $8.31 \times 10^{-3}$ | $1.59 \times 10^{-4}$ | ... | ... | ... |
| $\alpha_X$ | $\tau = -0.535$ | $\tau = -0.28$ | $\tau = -0.47$ | $\tau = -0.484$ | $\tau = -0.383$ | $\tau = -0.377$ |
| | $1.24 \times 10^{-4}$ | $4.42 \times 10^{-2}$ | $7.63 \times 10^{-4}$ | $5.27 \times 10^{-4}$ | $6.07 \times 10^{-3}$ | $6.83 \times 10^{-3}$ |
| $\alpha_{OX} - \alpha_X$ | $\tau = 0.597$ | $\tau = 0.330$ | $\tau = 0.507$ | $\tau = 0.542$ | $\tau = 0.450$ | $\tau = 0.445$ |
| | $1.86 \times 10^{-5}$ | $1.80 \times 10^{-2}$ | $2.81 \times 10^{-4}$ | $1.04 \times 10^{-4}$ | $1.26 \times 10^{-3}$ | $1.42 \times 10^{-3}$ |

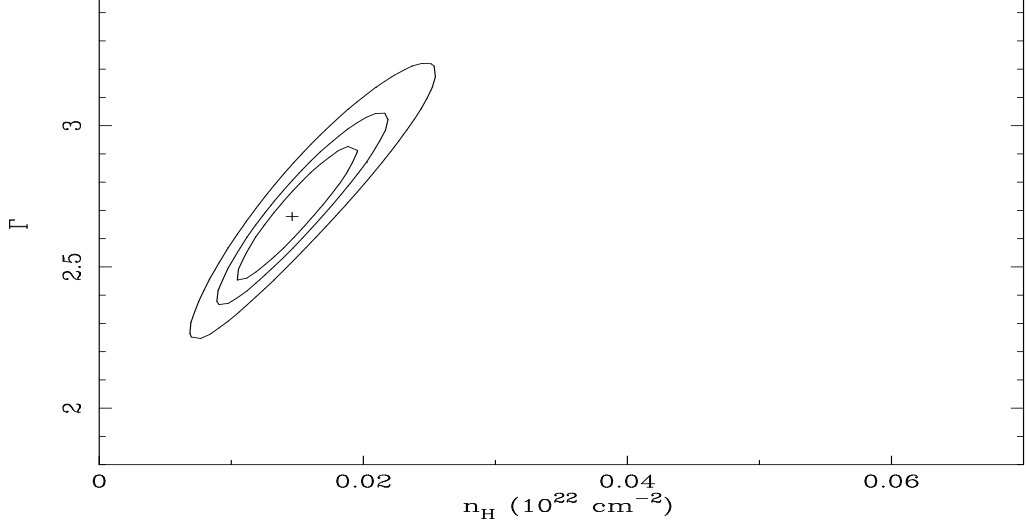

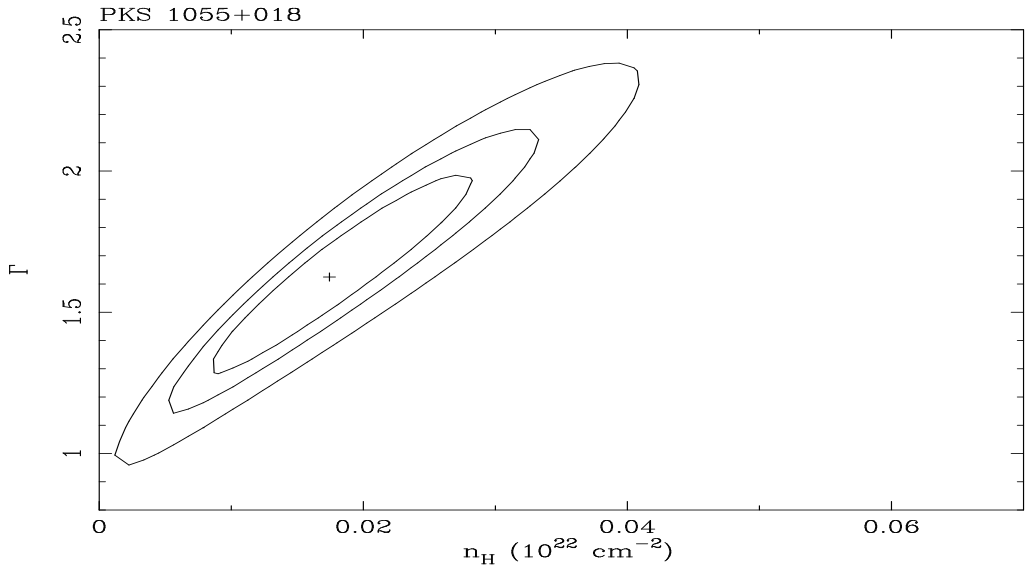

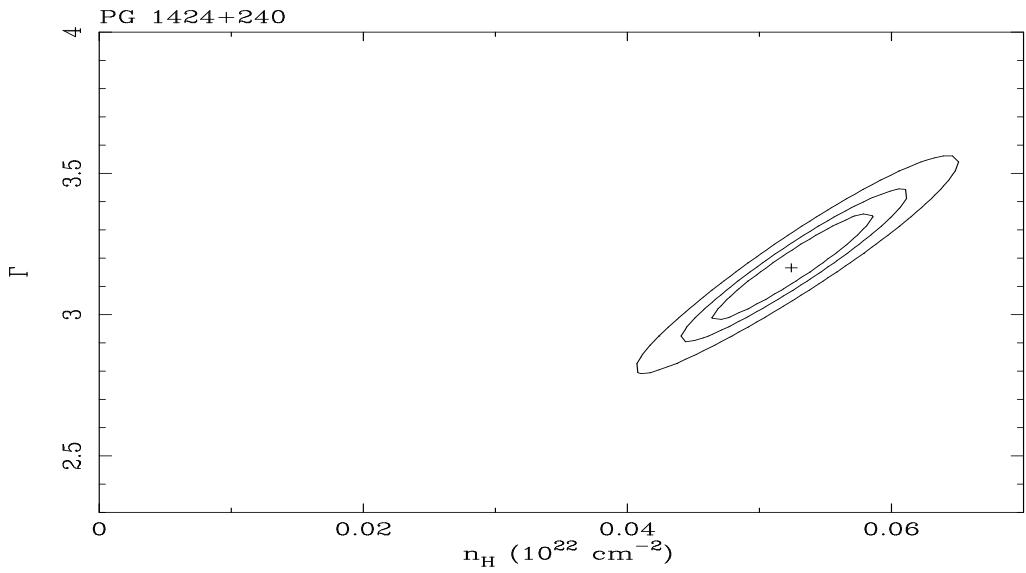

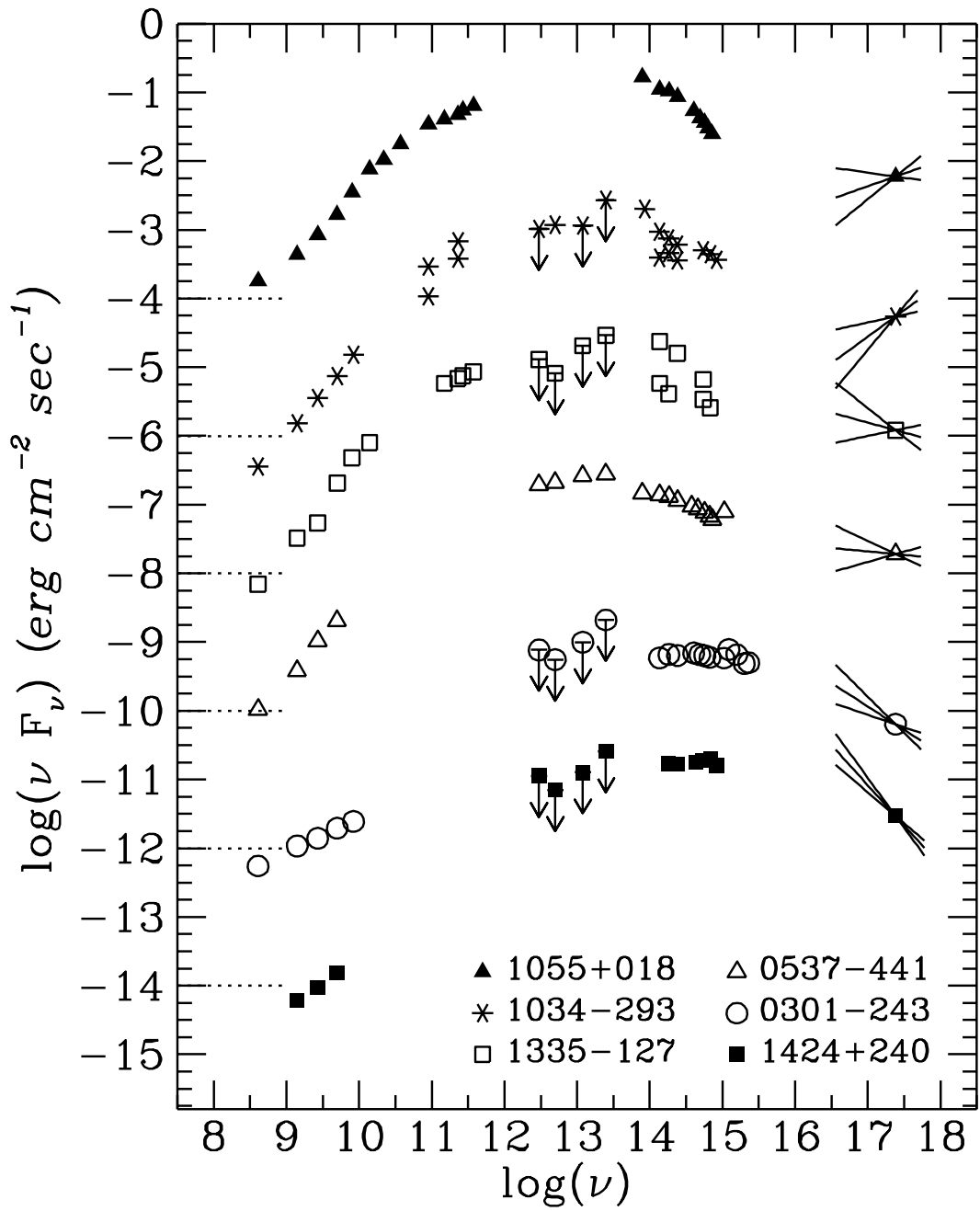

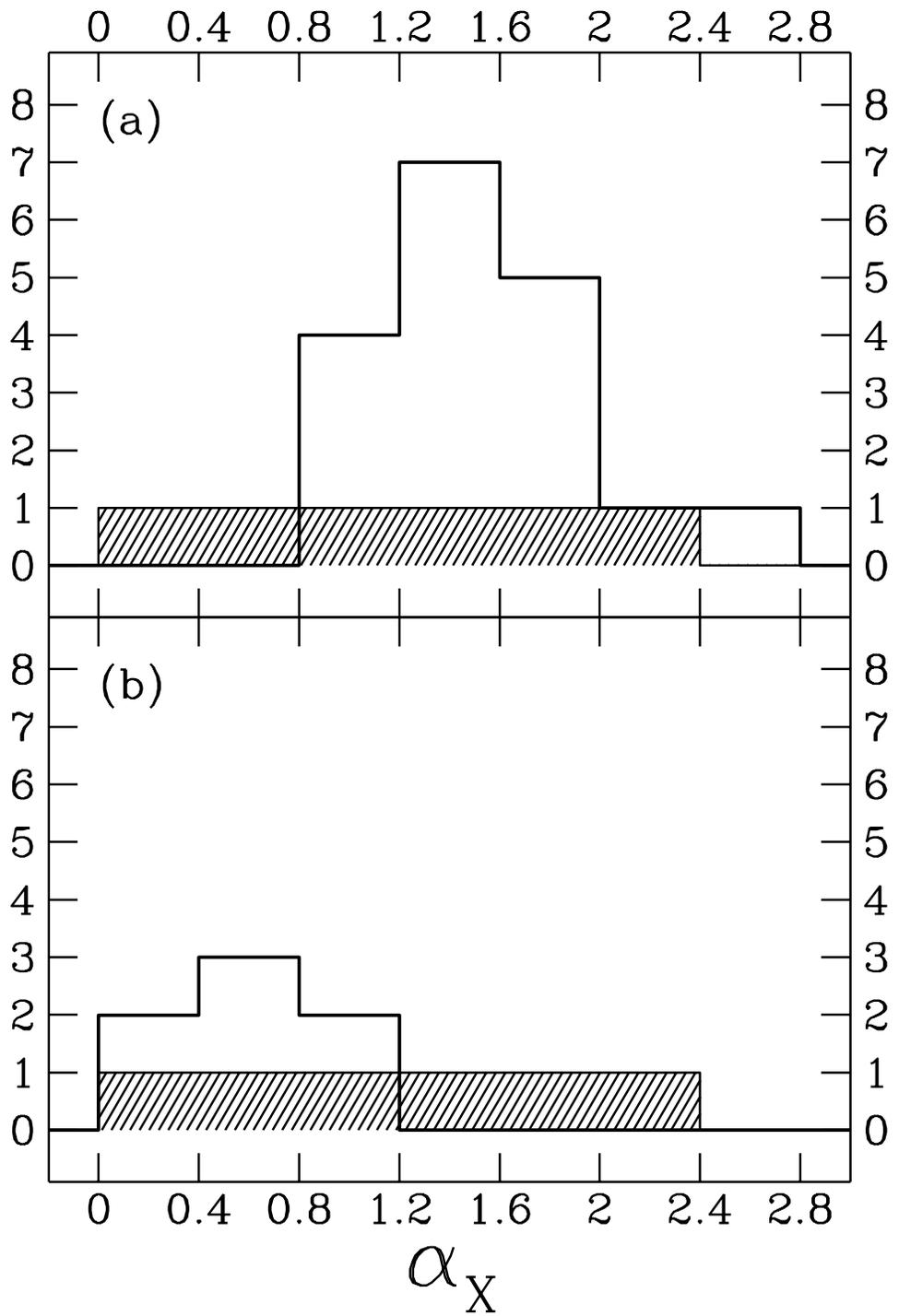

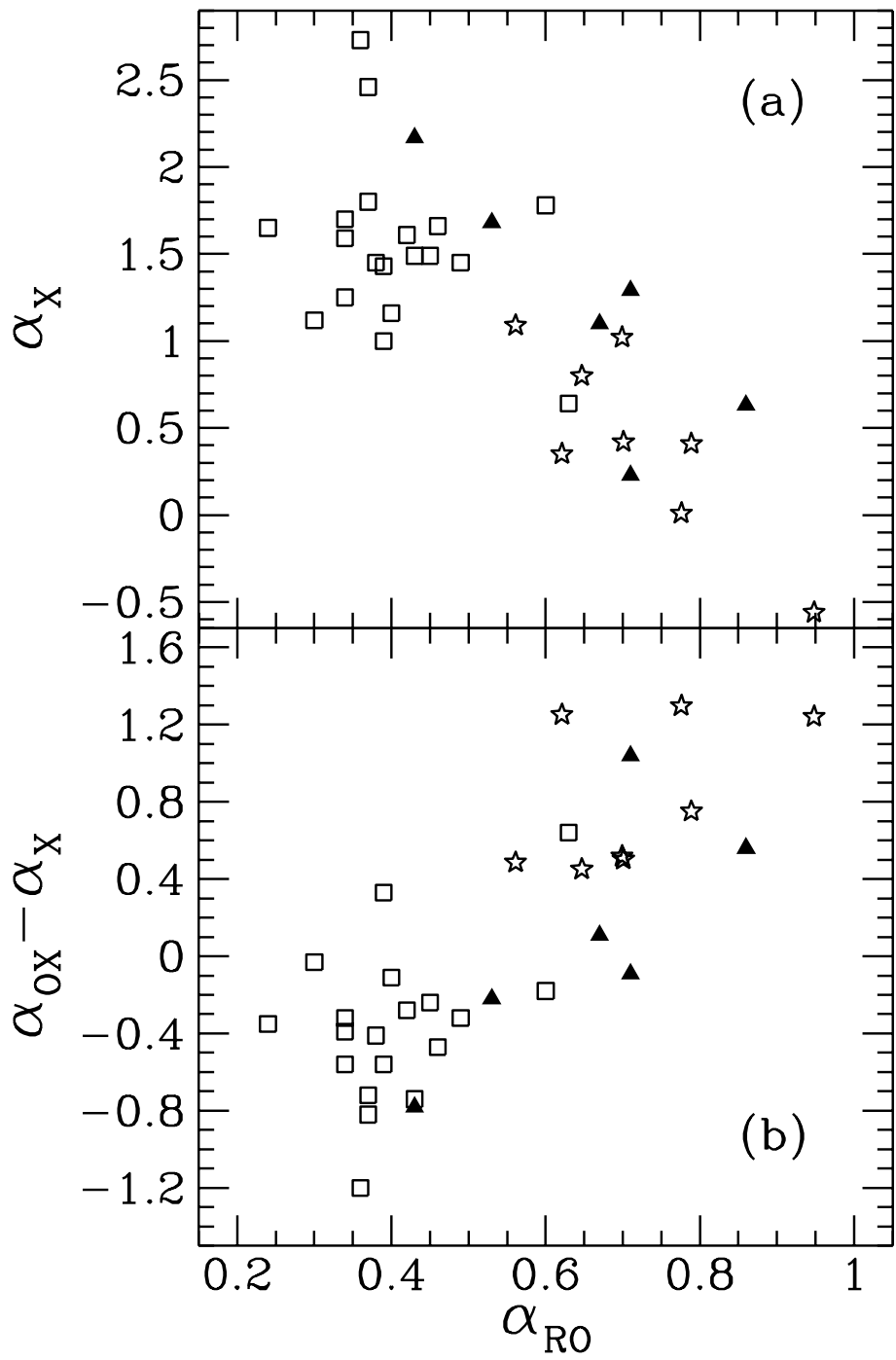